\documentclass[prd]{revtex4}

\usepackage{amsmath,amssymb}
\usepackage{graphicx}
\usepackage{pst-plot}
\usepackage{mathrsfs}
\usepackage{mathtools}
\usepackage{pst-func}

\DeclareMathAlphabet{\mathcal}{OMS}{cmsy}{m}{n}

\usepackage{titlesec}

\titlespacing\subsection{0pt}{12pt plus 2pt minus 2pt}{4pt plus 2pt minus 2pt}

\begin{document}

\title{Comment on the GHZ variant of Bell's theorem without inequalities}

\author{Joy Christian}

\email{jjc@bu.edu}

\affiliation{Einstein Centre for Local-Realistic Physics, Oxford OX2 6LB, United Kingdom}

\maketitle

\parindent 12pt

\baselineskip 11pt

In 1990, Greenberger, Horne, Shimony, and Zeilinger published a variant of Bell's theorem \cite{Bell-1964} without inequalities. Their argument published in \cite{GHZ} is now known as the GHZ theorem. It has since become one of the paradigmatic claims in the foundations of quantum mechanics, with several thousand citations to \cite{GHZ} to date.  The authors of \cite{GHZ} claim that the demonstration in \cite{EPR} by Einstein, Podolsky, and Rosen (EPR) of the incompleteness of the quantum mechanical description of physical reality is based on premisses that are inconsistent when applied to more than two particles in entangled quantum states. The authors claim that this inconsistency reveals that the EPR program for completing the quantum mechanical description of physical reality contradicts quantum mechanics even for the cases of perfect\break correlation among several spin-$\frac{1}{2}$ particles in entangled quantum states. Central to the authors' claim is the equation
\begin{equation}
{\mathscr A}(\phi'=\pi,\,\lambda)=+\,{\mathscr A}(\phi'=0,\,\lambda), \label{1n}
\end{equation}
where $\lambda$ represents hidden variables \cite{Bell-1964} and $\phi'$ represents an azimuthal angle in the geometry of the thought experiment considered by the authors (see Fig.~2 in \cite{GHZ}). By contrast, the equation consistent with the premisses of EPR should~be
\begin{equation}
{\mathscr A}(\phi'=\pi,\,\lambda)=-\,{\mathscr A}(\phi'=0,\,\lambda), \label{2m}
\end{equation}
which describes the perfect correlation between measurement results in the thought experiment considered by GHZ by extending the two-particle EPR-Bohm experiment \cite{Bohm-1951} to the system of several entangled fermionic particles. The authors' claim of inconsistency in the premisses of EPR rests entirely on the sign difference on the right-hand sides of these two equations. Unfortunately, the derivation of (\ref{1n}) in \cite{GHZ} is erroneous. The most transparent derivation of (\ref{1n}) is described by the authors in their endnote~15, giving credit to Mermin, which they state as: ``... just multiply the first three equations of (12) and compare the result with the fourth to conclude'' the above result (\ref{1n}). Unfortunately, this seemingly innocuous procedure amounts to applying the product rule illegitimately to eigenvalues of {\it non-commuting} operators in a hidden variable theory. The first three equations of (12) in \cite{GHZ}, namely, (12a) to (12c), are eigenvalues of mutually non-commuting operators, with the first part of the authors' procedure amounting to equating the product of the eigenvalues of these non-commuting operators with the eigenvalue of their product. However, since the pioneering work on hidden variable theories by von~Neumann \cite{vonNeumann} and Kochen and Specker \cite{Kochen}, the product rule is known to be applicable {\it only} to the eigenvalues of {\it commuting} operators. To see this, consider a spin-$\frac{1}{2}$ particle as an example. Since the Pauli operators $\sigma_x$ and $\sigma_y$ do not commute, {\it i.e.}, $\left[\sigma_x,\,\sigma_y\right]\not=0$, the eigenvalues of their product $\sigma_x\,\sigma_y$ are not equal to the products of the eigenvalues of $\sigma_x$ and $\sigma_y$. Indeed, the eigenvalues of $\sigma_x$ and $\sigma_y$ are $\pm1$, whereas the eigenvalues of $\sigma_x\,\sigma_y=i\sigma_z$ are $\pm\,i$, and thus not even real. Consequently, the product rule does not hold, because, clearly, $(\pm1)(\pm1)=\pm1\not=\pm\,i$. Note that what is asserted here by the inapplicability of the product rule is not that the eigenvalues cannot be assigned to all of $\sigma_x$, $\sigma_y$, and $i\sigma_x\,\sigma_y$ simultaneously as required by realism, but rather that the relationship among these eigenvalues {\it cannot be multiplicative} \cite{Kochen}, because $\sigma_x$ and $\sigma_y$ do not commute. Once this mistake in (\ref{1n}) is identified, it is not difficult to correct it, as I have shown in greater detail in Section~VIII of \cite{Begs}. This correction, however, leads to (\ref{2m}), which is the correct prediction of the perfect correlation within the GHZ version of the EPR-Bohm experiment. Thus, contrary to its popularity, the central claim of the GHZ theorem, namely, the alleged inconsistency in the very premisses of the EPR program for completing quantum mechanics \cite{EPR}, is incorrect.

\end{document}